\documentclass[a4paper,12pt]{article}

\usepackage{ifpdf}

\newif\ifpdf
\ifx\pdfoutput\undefined
  \pdffalse
\else
  \pdfoutput=1
  \pdftrue
\fi

\RequirePackage{xspace} %
\RequirePackage{subfigure} %
\RequirePackage[centertags]{amsmath} %
\RequirePackage{amssymb}
\RequirePackage{wrapfig} %
\RequirePackage{calc} %
\RequirePackage{ifthen}
\RequirePackage{tabularx} %
\RequirePackage{flafter} %
\RequirePackage{fancyhdr} %

\ifpdf
  \RequirePackage[pdftex]{color}%
  \RequirePackage{colortbl}%
  \RequirePackage{array}%
  \RequirePackage[pdftex]{graphicx}

  \RequirePackage[ pdftex, plainpages = false, pdfpagelabels,
                 pdfpagelayout = useoutlines,
                 bookmarks,
                 breaklinks = true,
                 linktocpage,
                 colorlinks = true,
                 linkcolor = blue,
                 urlcolor  = blue,
                 citecolor = blue,
                 anchorcolor = blue,
                 hyperindex = true,
                 hyperfigures
                 ]{hyperref}

\else
  \RequirePackage{color}
  \RequirePackage{colortbl}
   \RequirePackage{array}
  \RequirePackage[dvips]{graphicx}
  \RequirePackage{hyperref}
  \usepackage{rotating}
\fi

\usepackage{makeidx} 
\usepackage{setspace} 
\usepackage{rotating} 
\usepackage{ecltree}
\usepackage{epic}
\usepackage{supertabular}  
\usepackage{color}
\usepackage{exscale}
\usepackage{fontenc}
\usepackage{ifthen}
\usepackage{latexsym}
\usepackage{makeidx}
\usepackage{syntonly}
\usepackage{inputenc}
\usepackage{graphicx}
\usepackage{setspace}
\usepackage{caption2}
\usepackage[english]{babel}
\usepackage[square, comma,numbers,sort&compress]{natbib}
\usepackage{hypernat}
\usepackage{boxedminipage}
\usepackage{framed}
\usepackage{longtable}
\usepackage[all]{hypcap}    
\usepackage{algorithm2e}
\usepackage{algorithmic}
\usepackage{lscape}
\usepackage{pdflscape}
\usepackage[T1]{fontenc}
\usepackage{microtype}

\newcommand{\note}[1]{\marginpar[left]{\singlespace \tiny #1}}

\renewcommand{\sectionmark}[1]%
      {\markright{\thesection\ #1}} 

\renewcommand{\note}[1]{}

\doublespace

\DisableLigatures[f]{encoding = *, family = * }

\begin{document}
\begin{center}
{\Large Using the stress function in the flow of generalized Newtonian fluids through pipes and
slits}
\par\end{center}{\Large \par}

\begin{center}
Taha Sochi
\par\end{center}

\begin{center}
{\scriptsize University College London, Department of Physics \& Astronomy, Gower Street, London,
WC1E 6BT \\ Email: t.sochi@ucl.ac.uk.}
\par\end{center}

\begin{abstract}
\noindent We use a generic and general numerical method to obtain solutions for the flow of
generalized Newtonian fluids through circular pipes and plane slits. The method, which is simple
and robust can produce highly accurate solutions which virtually match any analytical solutions.
The method is based on employing the stress, as a function of the pipe radius or slit thickness
dimension, combined with the rate of strain function as represented by the fluid rheological
constitutive relation that correlates the rate of strain to stress. Nine types of generalized
Newtonian fluids are tested in this investigation and the solutions obtained from the generic
method are compared to the analytical solutions which are obtained from the
Weissenberg-Rabinowitsch-Mooney-Schofield method. Very good agreement was obtained in all the
investigated cases. All the required quantities of the flow which include local viscosity, rate of
strain, flow velocity profile and volumetric flow rate, as well as shear stress, can be obtained
from the generic method. This is an advantage as compared to some traditional methods which only
produce some of these quantities. The method is also superior to the numerical meshing techniques
which may be used for resolving the flow in these systems. The method is particularly useful when
analytical solutions are not available or when the available analytical solutions do not yield all
the flow parameters.

\vspace{0.3cm}


\par\end{abstract}

\begin{center}

\par\end{center}

\clearpage
\section{Introduction} \label{Introduction}

There are many applications for the flow of generalized Newtonian fluids in circular pipes and
plane slits. These include biological flow in living organisms and fluid shipping in transport and
process industries. Hence various analytical and numerical methods have been developed and used to
obtain the flow parameters which include the stress function, rate of shear strain, flow velocity
profile, local viscosity and volumetric flow rate as functions of the conduit geometry, pressure
drop and fluid rheology \cite{Skellandbook1967, BirdbookAH1987}.

A very simple numerical method, which is very natural to use, is to employ the stress function,
which is defined in terms of the conduit geometry and pressure drop, directly. The method is based
on a simple combination of the easily obtained stress function with the fluid rheology to obtain
the rate of strain as a function of the velocity-varying spatial dimension of the conduit, i.e. the
radius for the pipe flow and the thickness dimension for the slit flow. The flow velocity profile
and the volumetric flow rate can then be obtained from simple numerical integrations. The local
viscosity can also be obtained from the fluid rheological relation as soon as the stress and rate
of strain are obtained.

The method is very generic, as well as being more convenient to implement and use, and hence the
solutions obtained from it can be more reliable than the solutions obtained from more sophisticated
methods. It is also very general since it can be applied whenever the rate of strain can be
expressed as an explicit or implicit function of the shear stress; a relation that usually can be
obtained from the fluid rheological equation. Hence, the method can be used to solve almost all the
generalized Newtonian flow problems in pipes and slits since the stress function for these systems
can be easily obtained and the appropriate rheological equations can be readily expressed in the
desired form. The method is particularly useful when the flow parameters, or some of which, cannot
be obtained analytically. Hence, it can provide a good alternative to the classical methods for
solving these problems which are traditionally solved by employing rather complicated methods such
as numerical discretization techniques like finite element and finite difference.

In section \S\ \ref{Method} we give a general description of the method and its theoretical
justification as well as the type of flow systems to which the method applies and the restrictions
and assumptions that should be observed in its application. In section \S\ \ref{Implementation} we
discuss practical issues about the implementation of the method. We also present a sample of the
volumetric flow rate solutions that were obtained from the method with comparison to similar
results obtained from the Weissenberg-Rabinowitsch-Mooney-Schofield (WRMS) method. The paper is
ended in section \S\ \ref{Conclusions} with general discussions and summary of the main objectives
and accomplishments of this investigation.

\clearpage
\section{Method} \label{Method}

We assume an incompressible, laminar, isothermal, steady, pressure-driven, fully-developed flow of
purely-viscous, time-independent fluids that are properly characterized by the following
generalized Newtonian fluid model

\begin{equation}\label{GNF}
\tau = \mu \gamma
\end{equation}
where the shear viscosity, $\mu$, and shear stress, $\tau$, depend only on the contemporary rate of
strain, $\gamma$, with no memory for the fluid of its deformation history. The effects of external
body forces, like gravitational attraction or electromagnetic interaction, as well as the edge
effects at the entry and exit zones of the conduit are assumed insignificant. Dependencies on
physical factors like temperature, which are not related to deformation, are also ignored assuming
fixed conditions or negligible contribution from these factors. The flow is also assumed to be in
shear mode with no significant extensional contributions.

As for the boundary conditions, a no-slip at the conduit wall is assumed and hence a zero velocity
condition at the fluid-solid interface is maintained. For the investigated types of rheology and
flow system, the flow velocity profile has a stationary derivative point at the symmetry center
line of the pipe and symmetry center plane of the slit which means zero stress and rate of strain
at these loci. For viscoplastic fluids, these stationary zones extend to include all the points at
the forefront of the flow profile whose stress falls below the fluid yield stress.

Concerning the type of conduit, we use circular pipe geometry with length $L$ and radius $r$, and
plane slit geometry with length $L$, width $W$ and thickness $2B$ where the latter dimension is the
smallest of the three. A pressure drop $\Delta p$ is imposed along the conduit length dimension
which defines the flow direction. The pipe is assumed straight with a cross sectional area that is
uniform in shape and size while the slit is assumed straight, long and thin with a uniform cross
section. For both types of conduit, rigid mechanical properties of the conduit wall are assumed and
hence the conduit wall is not deformable under the considered range of pressure drop. It is also
assumed that the slit is positioned symmetrically in its thickness dimension, $z$, with respect to
the plane $z=0$.

The generic method is based on the proportionality between the stress and the spatial coordinate in
the flow profile dimension, $s$, which stands for the radius $r$ in the case of pipe and for the
thickness dimension $z$ in the case of slit. Since the stress $\tau$ as a function of $s$ is known
from the above mentioned proportionality, then the rate of strain, $\gamma$, can be easily obtained
from the fluid rheological constitutive relation which correlates the rate of shear strain to the
shear stress as long as it can be put in the form $\gamma=\gamma(\tau)$ where the dependency of
$\gamma$ on $\tau$ can be explicit or implicit. If $\gamma$ is an explicit function of $\tau$, as
it is the case for example for Ellis fluids (refer to Table \ref{GTable}), then $\gamma$ can be
obtained directly by a simple substitution in the rheological relation. If, on the other hand,
$\gamma$ is an implicit function of $\tau$, as it is the case for example for Cross fluids (refer
to Table \ref{GTable}), then $\gamma$ can be obtained numerically using a simple numerical solver
based for instance on a bisection method. In both cases, the obtained rate of strain as a function
of $s$ can be used to obtain the velocity profile and subsequently the volumetric flow rate by
consecutive numerical integrations.

To be more specific, the stress in the investigated flow systems of circular pipes and plane slits
is a function of the conduit geometry and pressure drop. The shear stress as a function of the
velocity-varying dimension $s$ can be obtained by several methods which can be found in many fluid
mechanics and rheology textbooks, e.g. \cite{Skellandbook1967, BirdbookAH1987, WhiteBook2002}, and
hence we are not going to give a formal derivation; instead we give the final results with a sketch
of how it can be obtained. From a simple force balance argument based on applying the Newton second
law of mechanics to the non-accelerating steady flow, where the normal force exerted on the conduit
cross section is balanced by the shear stress force, it can be established that the stress is
proportional to $s$ as summarized in the following equations for pipe and slit respectively

\begin{equation}
\tau=\frac{\tau_{R}}{R}r   \hspace{2cm}   {\rm and}   \hspace{2cm}   \tau=\frac{\tau_{B}}{B}z
\end{equation}
where $\tau_{R}$ and $\tau_{B}$ are, respectively, the pipe and slit wall shear stress as given by

\begin{equation}
\tau_{R}=\frac{R\Delta p}{2L}   \hspace{2cm}   {\rm and}   \hspace{2cm}   \tau_{B}=\frac{B\Delta
p}{L}
\end{equation}
where $R$ is the pipe radius, $B$ is the slit half thickness, and $L$ is the conduit length across
which a pressure drop $\Delta p$ is exerted.

Since the spatial dependence of the stress function is known, the rate of strain as a function of
$s$ can be obtained from the rheological relation expressed in the form
$\gamma\left(\tau(s)\right)$, as given for a sample of generalized Newtonian fluids in Table
\ref{GTable}, and hence $\gamma(s)$ is obtained. The obtained strain rate function is then
integrated numerically with respect to the spatial coordinate of the flow profile, $s$, to obtain
the flow velocity profile where the no-slip at wall condition provides an initial value for the
flow velocity, $v=0$, which is then incremented in moving from the wall to the center during the
integration process. The numerically obtained flow velocity is then integrated numerically with
respect to the conduit cross sectional area perpendicular to the flow direction to obtain the
volumetric flow rate. For viscoplastic fluids, the zero stress condition at the pipe center line
and slit center plane is extended to include all the regions at the forefront of the flow profile
whose shear stress falls below the fluid yield stress. In Table \ref{GTable} the rheological
constitutive relations for nine fluid models employed in this study are presented. For Carreau and
Cross models, $\gamma$ is given as an implicit function of $\tau$ and hence a simple numerical
solver like bisection is required to obtain $\gamma$ as a function of $\tau(s)$ and hence as a
function of $s$.


\renewcommand{\arraystretch}{1.6}

\begin{table} [!h]
\caption{The rate of shear strain, $\gamma$, as a function of shear stress, $\tau$, for the nine
fluid models used in this investigation \cite{Skellandbook1967, BirdbookAH1987, CarreaubookKC1997,
Tannerbook2000, OwensbookP2002}. For Carreau and Cross models, $\gamma$ is given as an implicit
function of $\tau$. The meaning of the symbols can be obtained from Nomenclature \S\
\ref{Nomenclature}. \label{GTable}}
\begin{center} 
{
\begin{tabular}{|l|l|}
\hline
Model & Rate of Shear Strain\tabularnewline
\hline
Newtonian & $\gamma=\frac{\tau}{\mu_{o}}$\tabularnewline
Power Law & $\gamma=\sqrt[n]{\frac{\tau}{k}}$\tabularnewline
Ellis & $\gamma=\frac{\tau}{\mu_{e}}\left[1+\left(\frac{\tau}{\tau_{h}}\right)^{\alpha-1}\right]$\tabularnewline
Ree-Eyring & $\gamma=\frac{\tau_{c}}{\mu_{r}}\sinh\left(\frac{\tau}{\tau_{c}}\right)$\tabularnewline
Carreau & $\gamma\left[\mu_{i}+\left(\mu_{0}-\mu_{i}\right)\left(1+\lambda^{2}\gamma^{2}\right)^{\left(n-1\right)/2}\right]=\tau$ \tabularnewline
Cross & $\gamma\left[\mu_{i}+\frac{\mu_{0}-\mu_{i}}{1+\lambda^{m}\gamma^{m}}\right]=\tau$\tabularnewline
Bingham & $\gamma=\frac{\tau-\tau_{0}}{C'}$\tabularnewline
Herschel-Bulkley & $\gamma=\sqrt[n]{\frac{\tau-\tau_{0}}{C}}$\tabularnewline
Casson & $\gamma=\frac{\left(\tau^{1/2}-\tau_{0}^{1/2}\right)^{2}}{K}$\tabularnewline
\hline
\end{tabular}
}
\end{center}
\end{table}

\renewcommand{\arraystretch}{1.2}


\clearpage
\section{Implementation, Results and Assessment} \label{Implementation}

The generic method, as described in the last section, was implemented in a computer code using
standard numerical integration and bisection solver techniques. The method was then employed to
obtain solutions for the flow of nine types of generalized Newtonian fluid through circular pipes
and plane slits. The nine types of fluid are: Newtonian, power law, Ellis, Ree-Eyring, Carreau,
Cross, Bingham, Herschel-Bulkley and Casson. The numerical results of the volumetric flow rate
which are obtained from the generic method using wide ranges of fluid and conduit parameters were
thoroughly compared to the results obtained from the WRMS analytical method.

A representative sample of the results obtained from the two methods are presented in Figures
\ref{QP} and \ref{QS} where the fluid and conduit parameters of these examples are given in Table
\ref{FDTable}. As seen in these examples, the solutions obtained from the generic method agree very
well with the WRMS analytical solutions. The minor departure between the two methods is due mainly
to the nature of the generic method as it heavily relies on numerical techniques, i.e. bisection
solvers and successive numerical integrations, which can accumulate numerical errors.

The big advantage of using the generic method in the flow problems through circular pipes and plane
slits is that it is very general as it can be applied to almost any generalized Newtonian fluid
model regardless of the complexity of its constitutive relation as long as the rheological relation
can be casted as an explicit or implicit function of the form $\gamma\left(\tau(s)\right)$.
Moreover, as the method only involves very simple and reliable numerical techniques, like bisection
solvers and numerical integration techniques, high accuracy of the solution can be achieved. It
also has no convergence difficulties under normal conditions, as it does not require matrix solvers
and similar sophisticated computational techniques, and hence it can be very reliable. Furthermore,
the method is very simple and hence it is easy and convenient to implement and use; which is an
advantage on its own plus being a further contributing factor to the reliability of the solutions.
Hence, the generic method can produce virtually exact solutions that match in their accuracy any
actual or potential analytical solutions even for the fluids with complex rheology. The generic
method can therefore offer a better alternative to the commonly used numerical methods which employ
more sophisticated computational techniques like finite difference.

Although the generic method is presented here as a numerical technique, it can also be applied
analytically where the stress function, $\tau(s)$, is substituted in the rheological relation to
obtain the rate of strain, $\gamma(s)$, which can be integrated successively to obtain the flow
velocity profile and volumetric flow rate. This can provide alternative analytical forms that can
be useful in some cases for verification or other purposes.

\newcommand{\Hs}      {\hspace{-0.1cm}} %
\newcommand{\CIF}     {\centering \includegraphics[width=1.85in]} %
\newcommand{\Vmin}    {\vspace{-0.2cm}} %


\begin{figure}
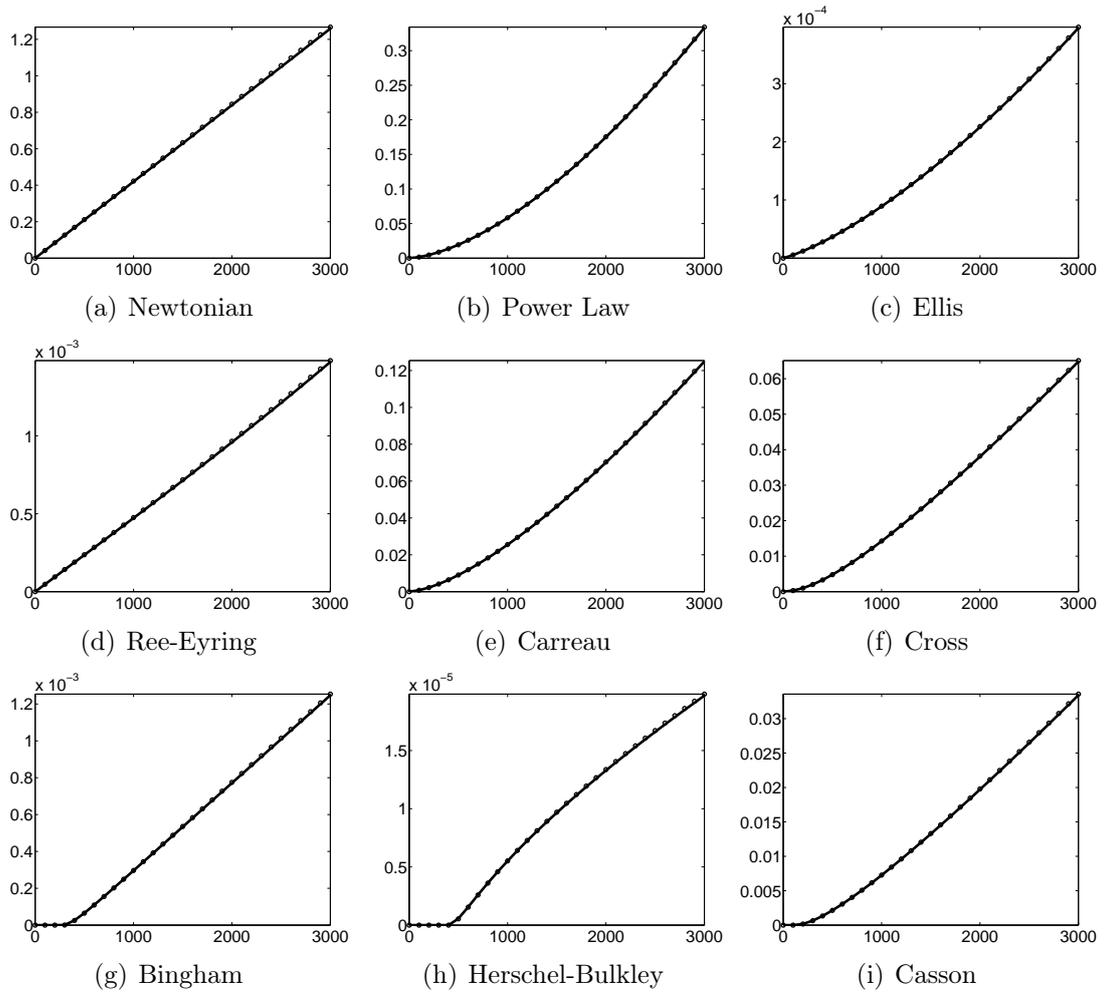

\centering %
\subfigure[Newtonian]%
{\begin{minipage}[b]{0.33\textwidth} \CIF {g/QP1}
\end{minipage}}
\Hs %
\subfigure[Power Law]%
{\begin{minipage}[b]{0.33\textwidth} \CIF {g/QP2}
\end{minipage}}
\Hs %
\subfigure[Ellis]%
{\begin{minipage}[b]{0.33\textwidth} \CIF {g/QP3}
\end{minipage}} \Vmin

%
\centering %
\subfigure[Ree-Eyring]%
{\begin{minipage}[b]{0.33\textwidth} \CIF {g/QP4}
\end{minipage}}
\Hs %
\subfigure[Carreau]%
{\begin{minipage}[b]{0.33\textwidth} \CIF {g/QP5}
\end{minipage}}
\Hs %
\subfigure[Cross]%
{\begin{minipage}[b]{0.33\textwidth} \CIF {g/QP6}
\end{minipage}}  \Vmin

%
\centering %
\subfigure[Bingham]%
{\begin{minipage}[b]{0.33\textwidth} \CIF {g/QP7}
\end{minipage}}
\Hs %
\subfigure[Herschel-Bulkley]%
{\begin{minipage}[b]{0.33\textwidth} \CIF {g/QP8}
\end{minipage}}
\Hs %
\subfigure[Casson]%
{\begin{minipage}[b]{0.33\textwidth} \CIF {g/QP9}
\end{minipage}}
\caption{Comparing the WRMS analytical solutions (solid line) with the numerical solutions from the
generic method (circles) of $Q$ in m$^3$.s$^{-1}$ (vertical axis) versus $\Delta p$ in Pa
(horizontal axis) for the flow of the nine fluid models in pipes. The pipe and fluid parameters are
given in Table \ref{FDTable}. \label{QP}}
\end{figure}


\begin{figure}
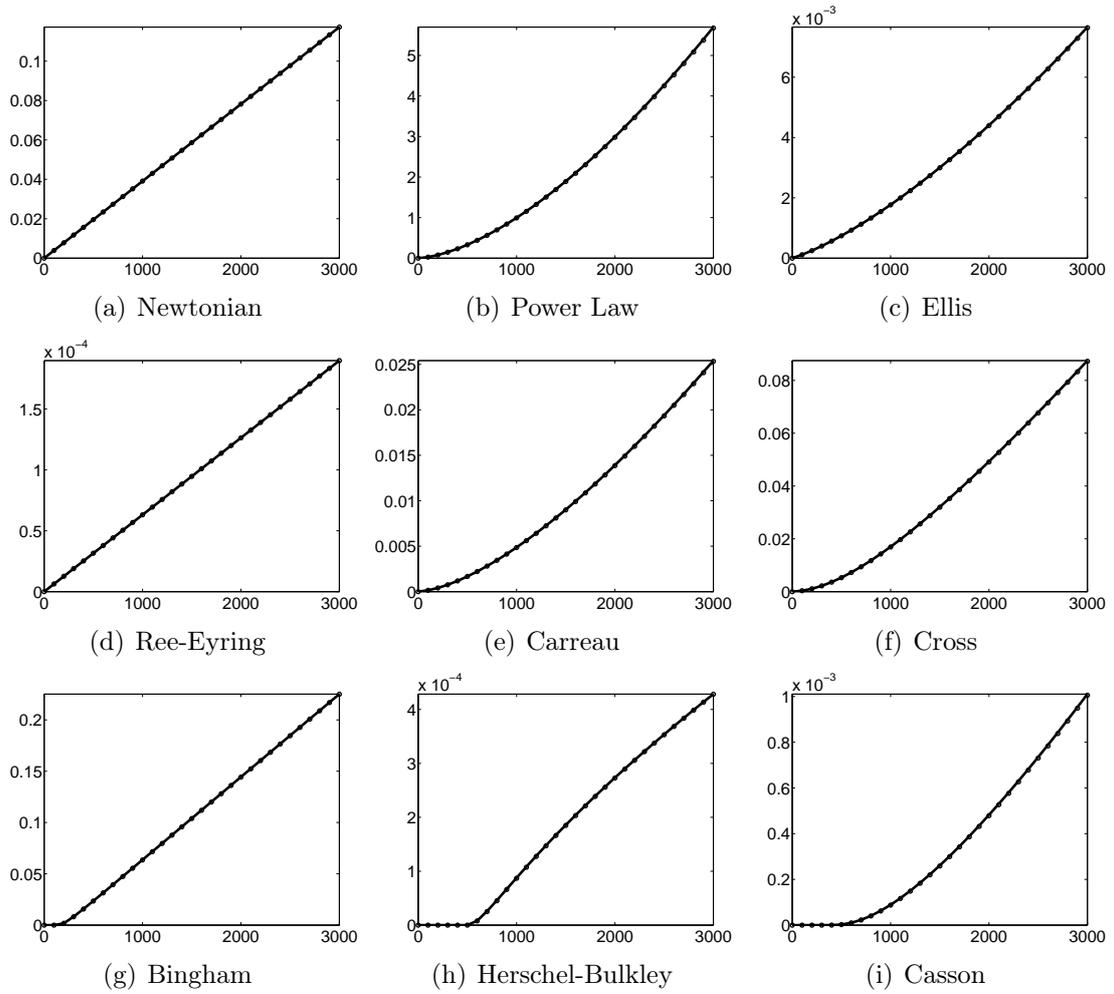

\centering %
\subfigure[Newtonian]%
{\begin{minipage}[b]{0.33\textwidth} \CIF {g/QS1}
\end{minipage}}
\Hs %
\subfigure[Power Law]%
{\begin{minipage}[b]{0.33\textwidth} \CIF {g/QS2}
\end{minipage}}
\Hs %
\subfigure[Ellis]%
{\begin{minipage}[b]{0.33\textwidth} \CIF {g/QS3}
\end{minipage}} \Vmin

%
\centering %
\subfigure[Ree-Eyring]%
{\begin{minipage}[b]{0.33\textwidth} \CIF {g/QS4}
\end{minipage}}
\Hs %
\subfigure[Carreau]%
{\begin{minipage}[b]{0.33\textwidth} \CIF {g/QS5}
\end{minipage}}
\Hs %
\subfigure[Cross]%
{\begin{minipage}[b]{0.33\textwidth} \CIF {g/QS6}
\end{minipage}}  \Vmin

%
\centering %
\subfigure[Bingham]%
{\begin{minipage}[b]{0.33\textwidth} \CIF {g/QS7}
\end{minipage}}
\Hs %
\subfigure[Herschel-Bulkley]%
{\begin{minipage}[b]{0.33\textwidth} \CIF {g/QS8}
\end{minipage}}
\Hs %
\subfigure[Casson]%
{\begin{minipage}[b]{0.33\textwidth} \CIF {g/QS9}
\end{minipage}}
\caption{Comparing the WRMS analytical solutions (solid line) with the numerical solutions from the
generic method (circles) of $Q$ in m$^3$.s$^{-1}$ (vertical axis) versus $\Delta p$ in Pa
(horizontal axis) for the flow of the nine fluid models in slits. The slit and fluid parameters are
given in Table \ref{FDTable}. In all these examples $W=1.0$~m. \label{QS}}
\end{figure}


\begin{table} [!h]
\caption{Fluid and conduit parameters for the examples of Figures \ref{QP} and \ref{QS}. The `R'
column applies to pipes and the `B' column applies to slits; the other columns are common to both.
SI units apply to all dimensional quantities as given in Nomenclature \S\ \ref{Nomenclature}, and
`HB' stands for Herschel-Bulkley. \label{FDTable}}
\begin{center} \vspace{-0.4cm}
{\small
\begin{tabular}{|l|l|c|c|c|}
 \hline
Model & Fluid Properties & $R$ & B & $L$\tabularnewline
 \hline
Newtonian & $\mu_{o}=0.013$ & 0.05 & 0.007 & 0.45\tabularnewline
 \hline
Power Law & $k=0.023,\,\,\,\,\,\,\,\,\,\, n=0.63$ & 0.015 & 0.008 & 0.55\tabularnewline
 \hline
Ellis & $\mu_{e}=0.026,\,\,\,\,\,\,\,\,\,\,\tau_{h}=8,\,\,\,\,\,\,\,\,\,\,\alpha=1.6$ & 0.005 & 0.002 & 0.22\tabularnewline
 \hline
Ree-Eyring & $\mu_{r}=0.41,\,\,\,\,\,\,\,\,\,\,\tau_{c}=43$ & 0.03 & 0.004 & 1.65\tabularnewline
 \hline
Carreau & $\mu_{0}=0.37,\,\,\,\,\,\,\,\,\,\,\mu_{i}=0.0079,\,\,\,\,\,\,\,\,\,\,\lambda=0.65,\,\,\,\,\,\,\,\,\,\, n=0.62$ & 0.043 & 0.0086 & 0.95\tabularnewline
 \hline
Cross & $\mu_{0}=0.42,\,\,\,\,\,\,\,\,\,\,\mu_{i}=0.0093,\,\,\,\,\,\,\,\,\,\,\lambda=0.77,\,\,\,\,\,\,\,\,\,\, m=0.58$ & 0.032 & 0.01 & 1.26\tabularnewline
 \hline
Bingham & $C'=0.033,\,\,\,\,\,\,\,\,\,\,\tau_{0}=5.7$ & 0.01 & 0.01 & 0.25\tabularnewline
 \hline
HB & $C=0.042,\,\,\,\,\,\,\,\,\,\,\tau_{0}=7.9,\,\,\,\,\,\,\,\,\,\, n=1.34$ & 0.005 & 0.002 & 0.13\tabularnewline
 \hline
Casson & $K=0.71,\,\,\,\,\,\,\,\,\,\,\tau_{0}=3.2$ & 0.08 & 0.011 & 1.27\tabularnewline
 \hline
\end{tabular}
}
\end{center}
\end{table}


\clearpage
\section{Conclusions} \label{Conclusions}

In this study we presented a generic and general numerical method for finding the flow solutions of
generalized Newtonian fluids in one dimensional flow problems that can be applied easily to
circular pipes and plane slits. The method can be used to obtain all the required flow parameters
which include shear stress, local viscosity, shear rate, flow velocity profile and volumetric flow
rate.

Thorough comparisons were made between the results of the generic method and the results of the
analytical solutions obtained from the Weissenberg-Rabinowitsch-Mooney-Schofield method. In all
cases, the two methods produced virtually identical results considering the numerical errors
introduced by the heavy use of numerical methods, like numerical integration and bisection solvers,
in the generic method.

The generic method enjoys several advantages over the competing numerical methods like finite
element and finite difference. These advantages include ease and convenience to implement and use,
generality, reliability and accuracy. The method may also be useful to apply analytically in some
cases.

The generic method can be particularly useful when no analytical solutions can be obtained, or the
analytical solutions can provide only some of the flow parameters, e.g. volumetric flow rate, but
not others, e.g. flow velocity profile.

\clearpage
\section{Nomenclature}\label{Nomenclature}

\begin{supertabular}{ll}
$B$                     &   slit half thickness (m) \\
$C$                     &   viscosity coefficient in Herschel-Bulkley model (Pa.s$^{n}$) \\
$C'$                    &   viscosity coefficient in Bingham model (Pa.s) \\
$k$                     &   viscosity coefficient in power law model (Pa.s$^n$) \\
$K$                     &   viscosity coefficient in Casson model (Pa.s) \\
$L$                     &   conduit length (m) \\
$m$                     &   indicial parameter in Cross model \\
$n$                     &   flow behavior index in power law, Carreau and Herschel-Bulkley models \\
$\Delta p$              &   pressure drop across the conduit length (Pa) \\
$Q$                     &   volumetric flow rate (m$^{3}$.s$^{-1}$) \\
$r$                     &   radius (m) \\
$R$                     &   pipe radius (m) \\
$s$                     &   spatial coordinate that represents $r$ for pipe and $z$ for slit (m) \\
$v$                     &   fluid velocity in the flow direction (m.s$^{-1}$) \\
$W$                     &   slit width (m) \\
$z$                     &   spatial coordinate of slit thickness dimension (m) \\
\\
$\alpha$                &   indicial parameter in Ellis model \\
$\gamma$                &   rate of shear strain (s$^{-1}$) \\
$\lambda$               &   characteristic time constant in Carreau and Cross models (s) \\
$\mu$                   &   fluid shear viscosity (Pa.s) \\
$\mu_{0}$               &   zero-shear viscosity in Carreau and Cross models (Pa.s) \\
$\mu_e$                 &   low-shear viscosity in Ellis model (Pa.s) \\
$\mu_{i}$               &   infinite-shear viscosity in Carreau and Cross models (Pa.s) \\
$\mu_{o}$               &   Newtonian viscosity (Pa.s) \\
$\mu_{r}$               &   characteristic viscosity in Ree-Eyring model (Pa.s) \\
$\tau$                  &   shear stress (Pa) \\
$\tau_0$                &   yield stress in Bingham, Herschel-Bulkley and Casson models (Pa) \\
$\tau_B$                &   shear stress at slit wall (Pa) \\
$\tau_c$                &   characteristic shear stress in Ree-Eyring model (Pa) \\
$\tau_{h}$              &   shear stress when viscosity equals $\frac{\mu_e}{2}$ in Ellis model (Pa) \\
$\tau_R$                &   shear stress at pipe wall (Pa) \\
\end{supertabular}

\clearpage
\phantomsection \addcontentsline{toc}{section}{References} %
\bibliographystyle{unsrt}

\end{document}